\documentclass[runningheads]{llncs}
\usepackage{graphicx}
\usepackage[utf8]{inputenc}
\usepackage{url}
\usepackage{hyperref}

\begin{document}
\title{Sharing and Learning Alloy on the Web}
%
%
\author{Nuno Macedo\inst{1,2} \and
Alcino Cunha\inst{1,2} \and
José Pereira\inst{2} \and
Renato Carvalho\inst{1,2} \and
Ricardo Silva\inst{2}\and
Ana C. R. Paiva\inst{1,3} \and
Miguel Sozinho Ramalho\inst{1,3} \and
Daniel Silva\inst{3}
}
\authorrunning{N. Macedo et al.}
%
\institute{INESC TEC
\and
Universidade do Minho 
\and
Universidade do Porto
}
\maketitle              
\begin{abstract}
We present Alloy4Fun, a web application that enables online editing and sharing of Alloy models and instances, to be used mainly in an educational context. By introducing the notion of secret paragraphs and commands in the models, it also allows the distribution and automatic evaluation of simple specification challenges, a useful mechanism that enables students to learn relational logic at their own pace. Alloy4Fun stores all versions of shared and analyzed models, as well as derivation trees that depict how those models evolved over time: this wealth of information can be mined by researchers or tutors to identify, for example, learning breakdowns in the class or typical mistakes made by students and other Alloy users. A beta version of Alloy4Fun was already used in two formal methods courses, and we present some results of this preliminary evaluation.
\keywords{Teaching formal methods \and Alloy \and Web application}
\end{abstract}

\section{Introduction}

Alloy~\cite{alloy} is a popular formal specification language and toolkit to describe and reason about a software design. It is also taught in several undergraduate and graduate courses in formal methods\footnote{A (bit outdated) list of courses using Alloy and papers reporting on the experience of teaching with Alloy can be found in its official website at \url{www.alloytools.org}.}, including our own graduate courses at University of Minho and University of Porto, in Portugal. One of the reasons for this popularity is the support for automated analysis provided by the Alloy Analyzer, an easy to download and install self-contained executable written in Java. The Analyzer also allows instances (either witness scenarios or counter-examples) to be graphically depicted using user-customized themes, a popular feature both for experienced users and students.

Despite such streamlined toolkit, over the many years we taught and researched with Alloy we identified some missing features and functionalities that could further ease its adoption, in particular in an educational context:
\begin{itemize}
    \item A straightforward mechanism to share simple Alloy models and instances. 
    \item Some auto-grading functionality or online judge system for students to automatically check the correctness of their exercise resolutions. 
    \item A mechanism to quickly obtain insight on how students use the language, namely identify typical mistakes or learning breakdowns in the class. 
\end{itemize}

To address these limitations 
 we 
developed Alloy4Fun, a web application that enables online editing and sharing of Alloy models and instances, including simple specification challenges in the form of duels where students attempt to discover a secret teacher's specification. 


In this paper we present Alloy4Fun, starting with an overview of (and rationale for) its current features in Section~\ref{features}. Then, in Section~\ref{evaluation} we  briefly describe its implementation and some  results from the evaluation of a beta version in two formal methods courses. Finally, in Section~\ref{conclusion} we conclude the paper and present some ideas for future work. 
Knowledge of Alloy is not required to understand the paper, but can help better appreciate some of the features of Alloy4Fun.

\section{Alloy4Fun overview}
\label{features}

The core of Alloy4Fun, mimics in a web application, the main features of the standalone Alloy Analyzer. 
After accessing \url{alloy4fun.inesctec.pt} the user gets an empty online editor (with syntax highlighting) where Alloy models can be written. 
A button allows analysis commands in the model to be executed, either \texttt{run} commands to get witness scenarios or \texttt{check} commands to find counter-examples to properties. If found, these instances are depicted bellow the editor as graphs that, likewise in Analyzer, can be customized with user-defined themes (including signature projection). 
Besides these core functionalities, Alloy4fun tries to improve some aspects of the Analyzer, which are described in the sequel.

\begin{figure}
    \centering
    \includegraphics[width=9cm]{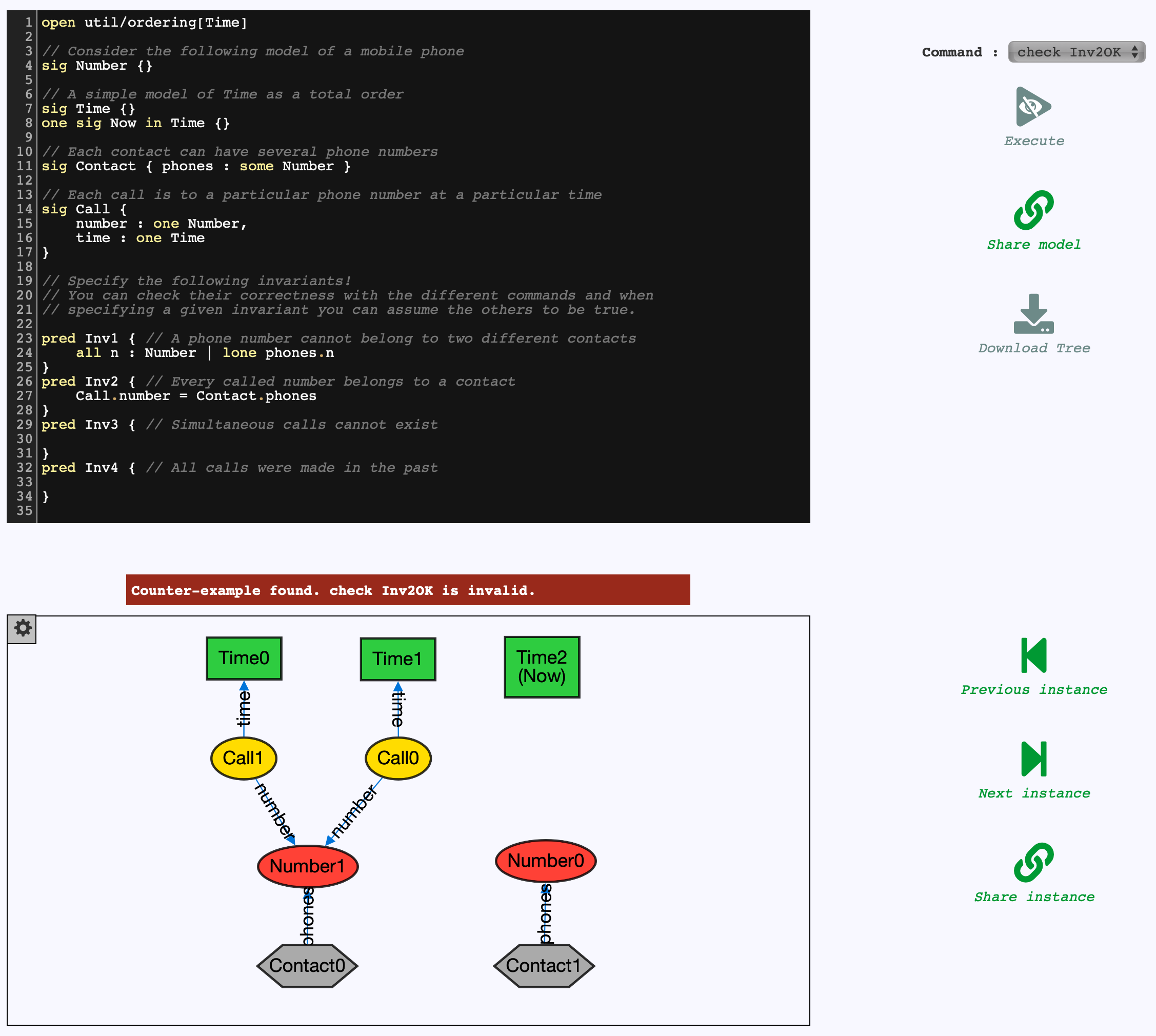}
    \caption{A failed attempt to solve a challenge}
    \label{fig:example}
\end{figure}

\textbf{Instance visualization and navigation.}
Unlike in the Analyzer, some of the most frequent theme customizations (like changing colors or shapes) can be performed quickly through a right-click menu on atoms or edges. Different layout algorithms can be selected to automatically organize nodes, which the user can manually modify. Unlike in the Analyzer, these positions are preserved between the frames of projected instances. Figure~\ref{fig:example} shows a screen capture of Alloy4Fun where a counter-example is being depicted with a user-defined theme. In addition, besides navigating to the next instance, the user can also re-visit previous instances, unlike in the Analyzer.

\textbf{Sharing models and instances.}
Currently, when someone wants to share or reference a model or instance, it is necessary to copy the model or screen captures, or setting up a web-site or repository. This rapidly becomes tedious and time consuming, in particular for tutors of large classes. Alloy4Fun provides the ability to easily share models and instances. After pressing the ``share model'' button a \textit{permalink} is generated, that can later be used to access the model. Any theme defined by the user is also preserved when sharing, thus allowing instances of shared models to be depicted as intended. Instances can also be shared via \textit{permalinks}. The theme and layout of the depicted atoms and relations at the time of sharing are preserved. This is a very handy feature since, likewise in the Analyzer, the positioning of atoms by the automatic layout mechanism is often not ideal, requiring manually rearrangement for better comprehension.

\textbf{Anonymous interaction.}
In Alloy4Fun there are no user accounts nor means to recover the \textit{permalinks} of previously shared models and instances. The user is responsible for keeping track of his \textit{permalinks} using some external mechanism, and Alloy4Fun provides a ``copy to clipboard'' button that helps with this task. The anonymity, namely the absence of user accounts, was a design choice in order to keep the interaction with the web application as simple as possible, to maximize user exposure, and also to avoid dealing with privacy and security issues, namely the hassle of storing and managing user credentials.

\textbf{Auto-grading.}
Although Alloy's specification language has very neat and simple syntax and semantics, many students struggle with its declarative nature, in particular those used to procedural programming~\cite{BoyattSinclair}. 
One way to overcome this difficulty is by independently solving exercises proposed by tutors, but even with automated analysis it is often difficult for students to assess whether they reached the correct answer, and tutors are required to inspect the solutions (something not scalable for large classes). 
These problems could be mitigated with an auto-grader, allowing students to solve exercises at their own pace and without the constant need for face-to-face time with tutors. In recent years, auto-graders and online judges have become widely popular for learning how to program \cite{6623681}, and we believe this success could be replicated in the learning of formal methods in general, and Alloy in particular.

With this in mind, the user in Alloy4Fun has the ability to mark any paragraph of a model as \emph{secret}. In Alloy a paragraph is either a signature (and its fields) declaration, a fact with a constraint that is assumed to hold, an assertion to be checked, an auxiliary predicate or function, or a \texttt{run} or \texttt{check} command. To mark a paragraph as secret, the special comment \texttt{//SECRET} must be added immediately before. When sharing a model with secret paragraphs two \textit{permalinks} are generated: a private one that, when accessed, reveals the full model, including secrets; and a public one that, when accessed, only shows public paragraphs, but internally still considers the secret in analyses and still allows the execution of secret commands (whose names are public). 
Using a comment instead of a new keyword to mark secret paragraphs ensures compatibility with Alloy's default syntax, allowing users to copy and paste models from Alloy4Fun to the stand-alone Analyzer, and vice versa.

This feature can be used to create simple specification challenges in the form of duels, where the user tries to reach a secret specification. Such models can have a public predicate that the user must fill-in, together with a secret \texttt{check} command that asserts (for a given scope) that such predicate is equivalent to the desired (hidden) specification. The model shown in Fig.~\ref{fig:example} was obtained precisely by accessing the public \textit{permalink}
of a model with 4 challenges (in this case, simple exercises where a natural language description of the desired specification is given for each of them)\footnote{The \textit{permalink} of this model and others can be found in a curated list of examples at \url{https://github.com/haslab/Alloy4Fun/wiki/Examples}.}. After filling the empty predicate (e.g., \texttt{Inv2}), a user can check whether it is a valid solution (e.g., secret command \texttt{Inv2OK} for \texttt{Inv2}), which will either return a ``no counter-example found'' message, meaning the challenge is solved, or a counter-example (as is the case in Fig.~\ref{fig:example}, where the counter-example shows that the specification \texttt{Inv2} is still not correct).

\textbf{Mining derivation trees.}
A possible way to gain insight about the students' learning process is to have access to their attempts at solving the proposed exercises, and tool support to mine this corpus for useful data \cite{8454297}.
Again, such feature would also be useful in a research context, and was one of the reasons that led Microsoft to develop the \url{www.rise4fun.com} web service, that allows researchers to easily deploy their tools on the web and collect human-tool interactions for posterior mining~\cite{rise4fun} (besides other advantages of web tools, like increased exposure, since the need for downloading and installing is eliminated, and promoting reliability given the large amount of test cases that can be collected for free). 
One of the most popular examples available via Rise4Fun, and the inspiration for developing Alloy4Fun, is \url{www.pex4fun.com}, a web-based educational gaming environment for learning programming, where students can engage in coding duels where they attempt to write code equivalent to a teacher's secret implementation~\cite{pex4fun}. Pex~\cite{pex}, an advanced white box test-generation tool, is used on the background to find inputs that show discrepancies between the student's code and the secret implementation. However, the interaction with the outcome of the tools is limited in Rise4Fun, which would prevent the implementation of key Alloy features like instance iteration and customization. This has led us to implement our own solution rather than integrate Alloy in this service.
    
Every shared model and instance is stored by Alloy4Fun in its database. However, to enable the proponents of challenges to mine the submissions for useful information, every command that was executed (and the respective result) and associated model are also stored. Moreover, for each model, the identifier of the model from which it derives and a time-stamp are also stored. This means that all the models that developed after accessing a shared \textit{permalink} end up forming a \emph{derivation tree}. In the case of a \textit{permalink} with challenges, a branch in this tree typically corresponds to an interactive session where one user / student is trying to solve the different challenges inside,   and can be analyzed to determine, for example, how many challenges where solved or how many attempts where needed to solve each one. Subsequent branches represent points where the a \emph{permalink} was generated for a model, and accessed multiple times. Alloy4Fun allows anyone in possession of the secret \textit{permalink} of a model to download the respective derivation tree in an easy to process JSON format. 

\section{Implementation and preliminary evaluation}
\label{evaluation}

We developed Alloy4Fun with Meteor, a full-stack isomorphic JavaScript framework for developing web applications based on Node.js. The client uses CodeMirror as text editor and the Cytoscape.js graph visualization library to depict instances. Models and instances are stored in a MongoDB document-oriented database at the server. To execute commands, we encapsulated the Alloy Analyzer in a RESTful web service implemented in Java. Seamless deployment of both the application and the service in a server is performed using Docker. All the Alloy4Fun code is open-source and available at \url{github.com/haslab/Alloy4Fun}.

We performed a preliminary evaluation of Alloy4Fun in the first semester of the 2018/19 academic year in two graduate formal methods courses at University of Minho (UM) and University of Porto (UP), in Portugal. The former teaches Alloy for 6 weeks and had 22 students enrolled, and the latter for 4 weeks and had 156 students enrolled. Both courses have one weakly lecture and one weakly lab session. 
The main goals of this evaluation were: 
1) to test a beta version of the application in a medium-sized audience, to identify bugs and possible design improvements; 
2) to determine whether students found Alloy4Fun useful as an auto-grader while learning Alloy.
Notice that the labs still relied mainly on the Alloy Analyzer for hands-on practice. Alloy4Fun use was not mandatory and was introduced to the students via a list of challenges with exercises for them to independently practice Alloy outside of the classes.

\begin{figure}
    \centering
    \includegraphics[width=10cm]{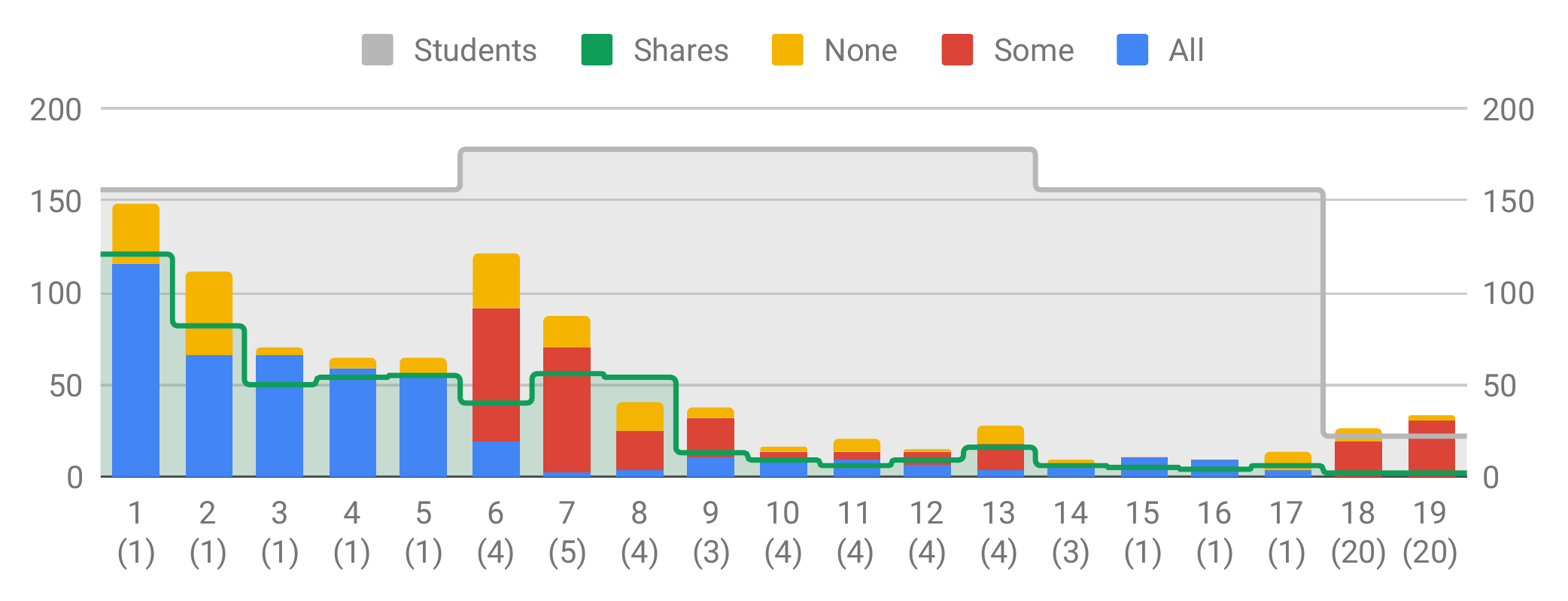}
    \caption{Results of the preliminary evaluation}
    \label{fig:evaluation}
\end{figure}


Regarding goal 1, we identified many bugs during this evaluation, including a major issue that caused a slowdown in the application that, unfortunately, rendered it almost unusable after a few weeks. All these bugs were fixed in the current version. We also identified several potential design improvements, most of them already implemented. For example, the beta version had a special ``lock'' comment to prevent the accidental edition of certain paragraphs. However, we noticed students rarely tried to change the model outside of the challenge predicates, and opted to remove this feature for simplicity and efficiency.

Regarding goal 2, we developed 19 different exercises containing challenges and provided them as shared models to students. These ranged from trivial exercises (e.g., asking to enforce simple inclusion dependencies or multiplicities), to more complex ones requiring the use of nested quantifiers or closures.  Figure~\ref{fig:evaluation} shows, for each shared model: the number of challenges inside (between parenthesis), the number of students that had access to the challenge, how many times \textit{permalinks} were generated by sharing, and how many solving sessions were recorded (in a cumulative stacked bar discriminating the number of sessions where all, some or no challenges were solved). As an example, exercise \#6 (containing 4 challenges) is the one shown in Fig.~\ref{fig:example}. 

Recall that a session is a branch in the derivation tree, typically recording the interaction of one student with Alloy4Fun while solving the challenges inside a model. Of course, some students might have multiple sessions recorded, as sometimes they did not solve all the challenges in a model in a single continuous session and accessed the shared \textit{permalink} several times, instead of generating a new \textit{permalink} of a partial resolution for later resuming the work. Even with this uncertainty, it is safe to say that almost all of the students used Alloy4Fun at least once and most of them used it several times, a very positive result in particular considering the aforementioned performance bug. 

\section{Concluding remarks and future work}
\label{conclusion}

We briefly presented Alloy4Fun, a web application for online editing and sharing Alloy models and instances, that also allows the distribution and auto-grading of simple specification challenges, by just marking some paragraphs (namely \texttt{check} commands) as secret. Its main intended use is in an educational context, and a preliminary evaluation of a beta version in two courses provided evidence that students found the auto-grading feature useful for learning Alloy. 
In the future, we intend to develop tools to simplify the mining of useful data from the derivation trees, possibly to be run server-side at the click of a button (with results visualized in the browser), and also incorporate an alternative instance visualizer more amenable for dynamic systems~\cite{anima}. We also intend to continue using Alloy4Fun in our formal methods courses in the upcoming years, and collect relevant data to conduct a detailed study on how students learn and use Alloy, namely identify which features of the language they find more challenging.

\section*{Acknowledgements}

We would like to thank Daniel Jackson for the helpful comments and suggestions about the design of Alloy4Fun, and also all the students that were beta testers. This work is financed by the ERDF – European Regional Development Fund through the Operational Programme for Competitiveness and Internationalisation - COMPETE 2020 Programme and by National Funds through the Portuguese funding agency, FCT - Fundação para a Ciência e a Tecnologia, within project POCI-01-0145-FEDER-016826. The second author was also supported by the FCT sabbatical grant with reference SFRH/BSAB/143106/2018.

%
%
%
\bibliographystyle{splncs04}
\bibliography{bibliography}

\end{document}